\documentclass[preprint2]{aastex}
\def\ltsima{$\; \buildrel < \over \sim \;$}
\def\simlt{\lower.5ex\hbox{\ltsima}}
\def\gtsima{$\; \buildrel > \over \sim \;$}
\def\simgt{\lower.5ex\hbox{\gtsima}}
\def\AA{$\; \buildrel \circ \over {\rm A}$}

\def\s{\ifmmode \widetilde \else \~\fi}
\def\={\overline}

\def\spose#1{\hbox to 0pt{#1\hss}}

\def\etal{{\it et al.\ }}

\def\lta{\mathrel{\spose{\lower 3pt\hbox{$\mathchar"218$}}
     \raise 2.0pt\hbox{$\mathchar"13C$}}}
\def\gta{\mathrel{\spose{\lower 3pt\hbox{$\mathchar"218$}}
     \raise 2.0pt\hbox{$\mathchar"13E$}}}
\def\Dt{\spose{\raise 1.5ex\hbox{\hskip3pt$\mathchar"201$}}}    % upper case
\def\dt{\spose{\raise 1.0ex\hbox{\hskip2pt$\mathchar"201$}}}    % lower case

\def\dotsfill{\leaders\hbox to 1em{\hss.\hss}\hfill}

\slugcomment{To appear in ApJ Letters}

\shorttitle{Lewis, Ibata \& Wyithe}
\shortauthors{Searching for MACHOs in Galaxy Clusters}
\begin{document}

\title{Searching for MACHOs in Galaxy Clusters}

%% Use \author, \affil, and the \and command to format
%% author and affiliation information.
%% Note that \email has replaced the old \authoremail command
%% from AASTeX v4.0. You can use \email to mark an email address
%% anywhere in the paper, not just in the front matter.
%% As in the title, you can use \\ to force line breaks.

\author{Geraint F. Lewis}
\affil{Anglo-Australian Observatory, P.O. Box 296, Epping, NSW 1710, Australia}
\email{gfl@aaoepp.aao.gov.au}

\author{Rodrigo A. Ibata}
\affil{Max-Plank Institut f\"ur Astronomie, 
K\"onigstuhl 17, D--69117 Heidelberg, Germany}
\email{ribata@mpia-hd.mpg.de}

\author{J. Stuart B. Wyithe}
\affil{School of Physics, University of Melbourne, Parkville, 
Vic 3052, Australia \& \\ 
Princeton University Observatory, Peyton Hall, Princeton, NJ 08544, USA}
\email{swyithe@astro.princeton.edu}

%% Notice that each of these authors has alternate affiliations, which
%% are identified by the \altaffilmark after each name.  Specify alternate
%% affiliation information with \altaffiltext, with one command per each
%% affiliation.

%% Mark off your abstract in the ``abstract'' environment. In the manuscript
%% style, abstract will output a Received/Accepted line after the
%% title and affiliation information. No date will appear since the author
%% does not have this information. The dates will be filled in by the
%% editorial office after submission.

\begin{abstract}
If  cluster dark  matter is  in the  form of  compact objects  it will
introduce  fluctuations  into the  light  curves  of distant  sources.
Current  searches  for  MACHOs   in  clusters  of  galaxies  focus  on
monitoring quasars  behind nearby  systems.  This paper  considers the
effect  of  such  a  compact  population  on  the  surface  brightness
distribution   of   giant  gravitationally   lensed   arcs.   As   the
microlensing  optical  depth is  significant  in  these clusters,  the
expected fluctuations are substantial  and are observable. Focusing on
the giant arc seen in Abell 370, we demonstrate that several `extreme'
events would  be visible  in a comparison  of HST observations  at two
epochs.   Utilizing  NGST,  long   term  monitoring  should  reveal  a
ubiquitous twinkling of brightness over the surface of the arcs.
\end{abstract}

\keywords{galaxies: clusters: general -- gravitational lensing -- dark matter}

\section{Introduction}
% PLAN:
% General
After many years of searching,  the nature of dark matter still eludes
us.    Over  the   last  decade,   however,  programs   searching  for
microlensing  variability  in  stars  in the  Magellanic  Clouds  have
revealed a population  of compact objects in front  of the LMC (Afonso
\etal\  1999; Alcock  \etal\ 2000a).   The EROS  team have  used their
observations to place  an upper limit of 10\% of  the mass fraction in
MACHOs in  the Galactic halo  (Lasserre \etal\ 2000), while  the MACHO
team  interprets their  data as  providing evidence  of  a significant
fraction,  perhaps 20\%.  The  recent discovery  of nearby  cold, high
proper motion, hydrogen atmosphere white dwarfs appears to support the
latter interpretation (Ibata \etal\ 2000; Hodgkin \etal\ 2000).  These
results are complicated by fact  that views towards the Galactic Bulge
appear  to be  overdense in  compact objects  (Alcock et  al.  2000b),
suggesting  that the  Galactic Halo  is  clumped (c.f.  Klypin et  al.
1999) and our view towards  the Magellanic Clouds simply represents an
underdense line-of-sight.  Currently,  therefore, the true fraction of
the Galactic Halo composed of compact objects is uncertain.

Galaxy clusters represent the largest bound accumulations of matter in
the Universe, and  a number of lines of  evidence, including kinematic
(Bird,  Dickey {\&}  Salpeter 1993),  X-ray (Henry  et al.   1993) and
gravitational lensing (Kneib et al.  1993) studies, indicate that they
must possess a substantial quantity  of dark matter.  In this paper we
address the  question of what would be  the observational consequences
if this dark matter were also composed of compact objects.

% Macho Search & Problems
An approach  to test  the nature of  cluster dark matter  was recently
suggested by  Walker \&  Ireland (1995) and  Tadros, Warren  \& Hewett
(1998; 2000).  As with compact  matter in the Galactic Halo, MACHOs in
clusters  are   expected  to  introduce   brightness  fluctuations  in
background sources, and  this is searched for in  a monitoring program
of $\sim600$ quasars behind  the Virgo cluster.  While initial results
are  promising, the  vicinity  of  the Virgo  cluster  means that  the
microlensing  optical  depth  is  small  $(\sim  0.001)$  and  induced
fluctuations are relatively rare.   The optical depth can be increased
by changing  the lensing geometry  and considering clusters  at higher
redshift.   With  this,  however,   it  is  difficult  to  identify  a
substantial  number of background  quasars without  greatly increasing
the  number  of  clusters  to  be  monitored.   The  problem  is  also
compounded  by the  difficulty of  distinguishing microlensing-induced
quasar variability from intrinsic mechanisms.

% Where we are going + current paper
Lewis  \&  Ibata  (2000)  addressed  the question  of  what  effect  a
cosmological distribution of compact objects would have on the surface
brightness   distributions  of   galaxies  at   $z<0.5$.    While  the
microlensing  optical depth is  quite small  $(\sigma\simlt0.04)$, the
resulting low-level  fluctuations, $\sim2\%$, are  observable. In this
paper we extend this analysis to the view of distant galaxies observed
through galaxy clusters  whose dark matter is composed  of MACHOs. The
ideal targets are the  giant gravitationally lensed arcs.  Focusing on
these systems  ensures that the  column density of matter  through the
cluster is substantial, resulting in an optical depth near unity; this
vastly  exceeds   `self-lensing',  with  an  optical   depth  of  only
$\sim10^{-5}$,  where  MACHOs microlens  galaxies  within the  cluster
(Gould 1995).   Potential sources abound,  as giant arcs  and extended
images have been identified in  more than thirty clusters (e.g Fort \&
Mellier 1994).  As the induced variability  will be searched  for on a
pixel-by-pixel basis,  each extended image presents a  large number of
resolution  elements which  can  be viewed  as  separate sources,  far
outweighing the potential number of background quasars.

% Plans
In Section~\ref{method} we outline  the numerical approach we adopt to
tackle this problem, focusing  on the simulation of microlensing light
curves.  As  an illustration  of the efficiency  of microlensing  as a
mechanism    for    introducing    variability,    we    present    in
section~\ref{a370} a case  study of the giant arc  seen in the cluster
Abell   370.    This   section   also  discusses   the   observational
considerations  for   a  monitoring   program  of  giant   arcs.   The
conclusions to this study are presented in Section~\ref{conclusions}.

\section{Method}\label{method}
In  examining microlensing of  sources in  the Magellanic  Clouds, the
goal is to identify the variation  in brightness of a single star as a
compact   object  passes   in  front   of  it.    Considering  instead
microlensing  towards more  distant sources,  individual stars  can no
longer be  resolved and induced  variations are identified  against an
unresolved stellar  background.  Termed `pixel-lensing'  (e.g.  Crotts
1992), the magnification of a star  results in the increase of flux in
an individual  resolution element, and hence appears  as a fluctuation
in  the  surface  brightness  of  a source.   Employing  an  efficient
technique called `Difference  Imaging', searches for such fluctuations
have recently proved fruitful  with the identification of microlensing
variability towards the  Galactic Bulge and M31 (Alcock  et al. 1999a,
1999b; Ansari et al. 1999).

In a  similar vein, observations of gravitationally  lensed giant arcs
do  not  resolve individual  stars,  rather  each  pixel over  an  arc
represents the  summed flux from a  population of stars.   As stars in
this  population are  magnified due  to gravitational  microlensing by
compact  objects  in  the  foreground cluster,  the  apparent  surface
brightness of the pixels will be seen to fluctuate. It is important to
note that at the substantial  optical depths considered in this paper,
each star in the population will be subject to microlensing and so the
overall  light  curve   seen  will  be  the  sum   of  the  individual
fluctuations  of  each star.  This  differs  substantially from  pixel
lensing in  the local group, where  the optical depth is  small and at
most only a small number of stars are significantly microlensed at any
instant.

In  simulating the  expected variability,  the following  procedure is
adopted; firstly,  for a particular set of  microlensing parameters, a
large  catalogue  of  light  curves,  $\mu_i(t)$,  are  generated.   A
population of stars,  $S_i$, is randomly selected such  that the total
luminosity  of the population  is $L_{POP}  = \Sigma\,  L(S_i)$.  Each
star is used to normalize a light curve selected from the catalogue to
produce a microlensed view of  the star. These individual light curves
are then combined  to produce a total microlensed  light curve for the
entire stellar population, $M(t)  = \Sigma\, \mu_i(t)\, L(S_i)$.  As a
consistency check,  the mean luminosity of the  resultant light curves
must      equal      $\langle\mu_{th}\rangle      L_{POP}$,      where
$\langle\mu_{th}\rangle$ is the mean  magnification of the arc induced
by  the  macrolensing  model   of  the  cluster.   Typically,  stellar
luminosity  functions increase  at fainter  magnitudes,  implying that
populations are  dominated, in number, by low  luminosity stars.  Such
faint  stars, even  if  substantially lensed,  will not  significantly
contribute to the  final light curve. Hence, for any  star for which $
\mu_{max}  L(S_i)  / \langle\mu_{th}\rangle  L_{POP}  < 0.005$,  where
$\mu_{max}$ is taken  to be $70$, the contribution  to the final light
curve  is  taken  as  uniform  and  equal  to  $\langle\mu_{th}\rangle
L(S_i)$; such an approach  reduces the required numerical calculations
in  constructing a light  curve.  As  with Lewis  \& Ibata  (2000), we
adopt the  stellar luminosity function as determined  by Jahrei\ss\ \&
Wielen (1997).
%

%Unlike microlensing  in the Galactic Halo, the  variability induced by
%gravitational microlensing  at large  optical depths exhibits  quite a
%complex  structure, with  extended quiescent  periods  punctuated with
%bursts  of  extreme, rapid  variability  (Kayser,  Refsdal \&  Stabell
%1986).  Several  reviews of the physics  of gravitational microlensing
%in this regime  have been presented (e.g.  Schneider,  Ehlers \& Falco
%1992; Wambsganss 1998)  and they will not be  reproduced here.  Due to
%the non-linear combination of the gravitational lensing effects of the
%individual  compact objects,  no analytical  description of  the light
%curves is available, and studies in this area have concentrated on the
%development  of numerical  techniques to  investigate  microlensing at
%large optical depths.
%

An  efficient method  of  generating high  optical depth  microlensing
light curves  for a point-like  source is employed (Witt  1993; Lewis,
Miralda-Escud\'{e},  Richardson \&  Wambsganss  1993).  This  approach
treats the  source trajectory as  a line of infinite  extent, ensuring
that it is numerically simple  to find all the microlensed images with
a contour following algorithm. This guarantees that the light curve is
not lacking flux due  to `undiscovered' images.  The efficiency allows
large samples of light curves to  be generated, the basis for a number
of extensive statistical studies (Lewis  \& Irwin 1995; Lewis \& Irwin
1996; Wyithe \& Webster 1999; Wyithe, Webster \& Turner 1999,2000).

While  the  approach  is  straight-forward, the  volume  of  potential
parameter space is very large and a full examination of this is beyond
the scope of this paper, and  a more extensive study will be presented
elsewhere.   As an  illustration of  the efficacy  of  microlensing on
introducing variability into  surface brightness distributions, we now
focus on the spectacular giant arc in Abell~370.

\section{Case Study: A370}\label{a370}
At a redshift of $z=0.37$, the rich galaxy cluster Abell~370 possesses
several  giant gravitationally  lensed  arcs (Soucail  et al.   1987).
Utilizing  their  spatial  extent,   detailed  modeling  of  the  mass
distribution  of Abell~370  has  been undertaken  (Kneib  et al  1993;
Abdelsalam et al.  1998), confirming the binary nature of the cluster.
The most  prominent giant arc is over  20 arcsecs in extent  and has a
spectroscopic redshift of $z=0.724$ (Soucail et al.  1988).  Examining
the models for  the mass distribution in Abell~370  in the vicinity of
this giant  arc, we choose a  fiducial model with an  optical depth of
$\sigma=0.6$  and  shear of  $\gamma=0.2$.   This  results  in a  mean
magnification of $\langle\mu_{th}\rangle=8.33$.
 
\subsection{Light Curves}\label{lightcurves}
A large sample of light curves was generated using the above parameter
set and combined  with the stellar luminosity function  to produce the
light   curves    of   an    entire   population   as    outlined   in
Section~\ref{method}. The source trajectory is aligned with the shear,
although simple  scaling relations  to other shear  orientations exist
(Lewis \& Irwin 1995). Each lens has a mass of $1 M_\odot$.

\begin{figure*}
\centerline{
\includegraphics[width=4in,angle=270]{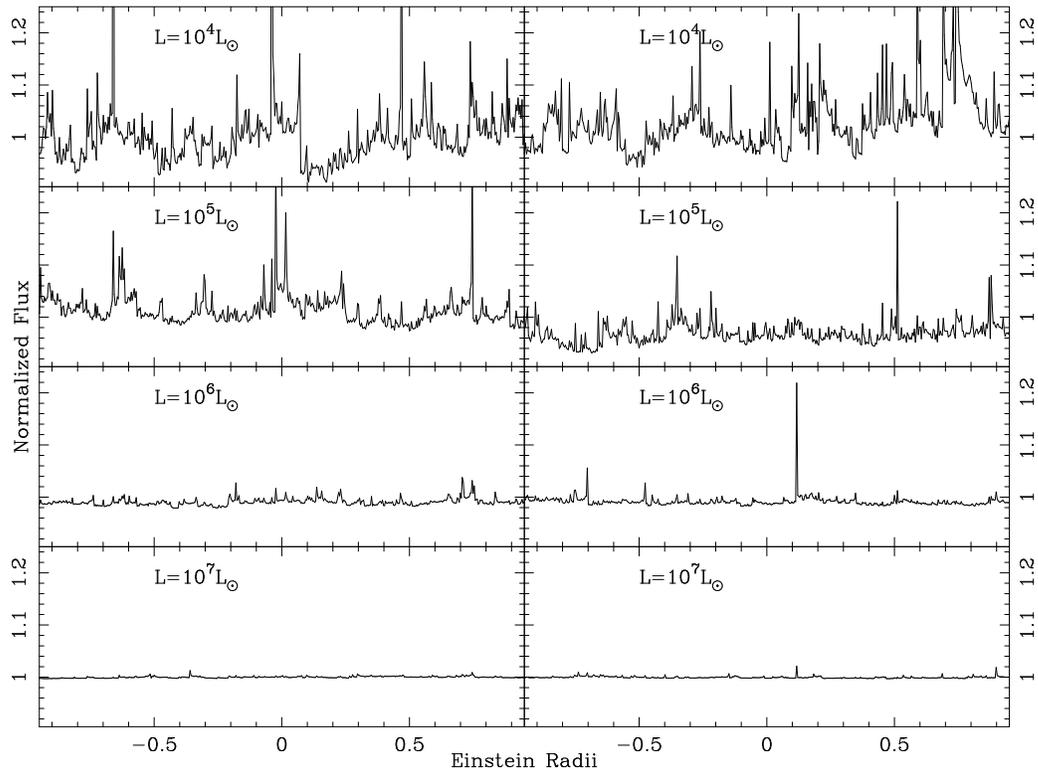}
}
\caption[]{Two sets  of examples of  light curves for  various stellar
populations. The  abscissa is  in units of  Einstein radii,  while the
ordinate presents the fluctuations with respect to the mean. Both sets
display the same general trends that are seen in all the light curves.
Namely, that  the lower luminosity  stellar populations show  the most
rapid  variability.  As  the  total luminosity  of  the population  is
increased,  even luminous  stars that  are strongly  lensed contribute
only small variations to the light curve.}
\label{fig1}
\end{figure*}

Several light  curves of these are presented  in Figure~\ref{fig1} for
stellar  populations   of  $L_{POP}   =  10^4$,  $10^5$,   $10^6$  and
$10^7L_\odot$; these numbers represent the intrinsic luminosity of the
population, the observed value  is $L_{Obs} = \langle \mu_{th} \rangle
L_{POP}$. The abscissa is in units  of Einstein radii for a solar mass
star; this is related to a  time scale, as explained below.  The light
curves  display   the  complex  variability  of   high  optical  depth
microlensing. At low optical depths, the light curves would consist of
solitary  events due to  individual microlenses.   While these  can be
used to directly infer microlensing parameters, they occur very rarely
and low optical depth light  curves are extremely quiescent.  With the
rich structure  seen in the light  curves of Figure~1,  the details of
the microlensing can be determined  in a statistical sense (e.g. Lewis
\& Irwin 1996).

In  examining the light  curves in  Figure~1 it  is apparent  that the
degree of variability drops as the total luminosity is increased, with
populations of $10^4L_\odot$ displaying rapid variations of 10-20\% on
scales much less  than an Einstein radius, while  the light curves for
populations of  $10^7L_\odot$ appear essentially flat  with only small
scale  variability.  This  is due  to the  fact that  even  a strongly
lensed, very luminous  star can contribute only a  small amount to the
total luminosity of a stellar population of $L_{POP}\sim10^7L_\odot$.

The  abscissa can  be normalized  into a  time scale  (Lewis  \& Ibata
2000); the crossing time of an Einstein radius of a Solar mass star is
9.2 $v_{1000}^{-1}  h^{-\frac{1}{2}}$ yrs (assuming  ${\rm \Omega_o=1,
\Lambda_o=0}$), where  $v_{1000}$ is the velocity  of the microlensing
masses  in  units  of  1000km/s.   When considering  a  population  of
microlensing objects of a different mass it is only this crossing time
that needs to be rescaled, with the degree of fluctuations in Figure~1
remaining the same.  Being simply a function of the square root of the
lensing  masses, if  the  dark  matter in  Abell~370  was composed  of
Jupiter   mass   compact    objects,   the
corresponding value would  be $\sim3.5 v_{1000}^{-1} h^{-\frac{1}{2}}$
months.

\begin{figure*}
\centerline{ 
\includegraphics[width=4in,angle=270]{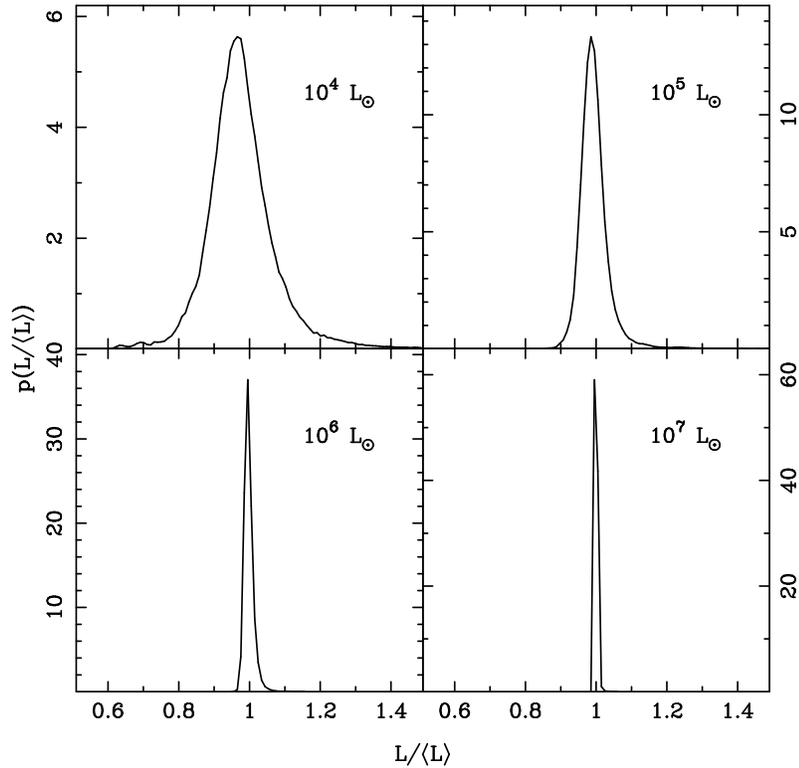} 
}
\caption[]{The   fluctuation   probability   distributions   for   the
populations  with intrinsic  luminosities  of $10^4,  10^5, 10^6$  and
$10^7L_\odot$. Note that the observed luminosity is $L_{Obs} = \langle
\mu_{th} \rangle  L_{POP}$ and the different ordinate  scaling on each
panel.    As   expected   from   the   light   curves   presented   in
Figure~\ref{fig1},  the  lower  luminosity  populations show  a  broad
distribution, which narrows as the total luminosity is increased.}
\label{fig2}
\end{figure*}

By combining the  light curves from the various  stellar samples it is
possible  to  generate a  probability  distribution  for the  expected
fluctuations (Figure~\ref{fig2}).  These reflect the features observed
in  Figure~\ref{fig1}, namely  that microlensing  of  lower luminosity
populations  results in  more substantial  fluctuations.  In  terms of
integrated  probability,  a   $10^4L_\odot$  population  will  exhibit
deviations from the  mean exceeding 5\% for half  of all observations.
A quarter of all observations will reveal fluctuations exceeding 10\%.
The corresponding values for the  more luminous populations are 2\% \&
3.5\% for $L_{POP}=10^5$, 0.6\% \& 1\% for $L_{POP}=10^6$ and 0.2\% \&
0.3\%  for   $L_{POP}=10^7$.   The  $L_{POP}=10^4L_\odot$   case  also
displays more extreme fluctuations, experiencing deviations of $>15\%$
for 10\% of observations, and $>29\%$ over 2\% of its light curve.

To detect the fluctuations expected for a population of cluster MACHOs
it  is  advantageous  to   target  stellar  populations  with  smaller
intrinsic luminosities.  Targeting fainter sources,  however, presents
observational   difficulties,    such   as   substantially   increased
integration times, a point we turn to in the next section.

\subsection{Observational Considerations}\label{observational}
The  study  of cosmological  microlensing  by  Lewis  \& Ibata  (2000)
examined  stellar luminosity  in  a  pixel covering  a  galaxy out  to
$z\sim0.5$.  Here we explore the possibility of using HST to constrain
the microlensing  variability in a  strongly lensed arc  of Abell~370.
For a  source redshift of  $z=0.724$, a spatially  resolved population
with  intrinsic U-band  luminosity  $L=10^7 L_\odot$,  magnified by  a
factor of  $8.3$, will give  rise to an  observed V-band flux  of $1.3
{\rm  photons/sec/m^2}$,  ($\Omega_o=1$,  $\Lambda_o=0$, and  $H_0=100
{\rm km/s/Mpc}$). The expected total  system efficiency of the new HST
instrument the ``Advanced Camera for Surveys'' is approximately $40$\%
at $6000$\AA,  so ${\rm 2.5  photons/sec}$ will be detected  per $10^7
L_\odot$  population.   The  sky  background  will  be  negligible  in
comparison, ${\rm  0.1 photons/sec}$ per resolution  element (the area
encompassed  by the  point spread  function)~\footnote{This  value was
derived  from  archival  WFPC2   ``Planetary  Camera''  data  on  this
cluster.}.   In  a  $10^4$~sec  exposure HST  will  therefore  measure
variability to 0.66\% accuracy; this would allow a $3\sigma$ detection
of 2\% variability events which according to the simulations of \S3.1,
occur with  0.25\% chance.  The  giant arc in Abell~370  subtends over
$\sim 1000$  HST spatial resolution elements and  $\sim 8$ variability
events would be detected in  a 2-epoch observation campaign if all the
cluster dark matter is in the  form of MACHOs.  Similarly, for a $10^6
L_\odot$ population, the above $10^4$~sec integration time would allow
3\%   variability   events   to   be   detected   at   the   $3\sigma$
level. Approximately  $\sim100$ such events  would be expected  in the
2-epoch dataset. Several exposures would  be required at each epoch to
correct for  cosmic rays and  to identify Poisson fluctuations  in the
surface photometry that could be mistaken for microlensing events.

Currently,  only  coarse  surface  photometry  of  the  giant  arc  in
Abell~370 has  been published, but its total  brightness suggests that
each HST resolution element corresponds to an average intrinsic source
luminosity of  $10^6\rightarrow10^7 L_\odot$.   HST images of  the arc
reveals that  it is very clumped (Smail  et al.  1996) and  a range of
source  luminosities are  available for  study. We  have begun  a more
detailed  modeling  of the  microlensing  in  Abell  370, taking  into
account both the variation  in the microlensing parameters and surface
brightness  over the  arc, with  the goal  of providing  a map  of the
variability signature.

To  detect  the  more  common  variability  in  populations  of  lower
luminosity  will   require  NGST.   With  this   instrument,  the  sky
background  is  expected  to  be  substantially  lower  than  on  HST:
$<10^{-1}$~photons/sec/resolution~element~\footnote{see
http://www.ngst.stsci.edu/sky/sky.html}  at  $1.1\mu  m$.  Assuming  a
$20$\% total system efficiency, a $10^5$~sec integration at two epochs
would allow  10\% variability events  to be detected at  the $3\sigma$
level  in a  population of  intrinsic luminosity  $10^4  L_\odot$.  As
discussed above,  variability of this  magnitude must be  very common,
occurring in  25\% of  the resolution elements  per epoch.   If MACHOs
make up a  substantial fraction of the dark  matter in clusters, giant
arcs will be seen to shimmer with NGST, the degree of shimmering being
dependent upon the luminosity of the population under consideration.

The  above  analysis  is  very  encouraging,  demonstrating  that  the
signature of  microlensing due  to cluster dark  matter could  soon be
detected  with   HST,  as  well  as   illustrating  that  microlensing
fluctuations may be ubiquitous in future observations with NGST.

\section{Conclusions}\label{conclusions}
If  dark matter  in  galaxy clusters  exists  in the  form of  compact
objects,  it  should  introduce  variability  on  background  sources.
Current searches for cluster  dark matter have focused upon monitoring
quasars behind the Virgo cluster.  Its close proximity, however, makes
Virgo a poor lens.

By considering  clusters at higher redshift,  the microlensing optical
depth increases,  although the potential number  of background quasars
that can be used as microlensing targets is reduced.  In this paper we
have  examined the microlensing  effect that  a population  of compact
dark matter objects would have on the surface brightness distributions
of  giant  luminous  arcs.   Focusing  upon  the  spectacular  arc  in
Abell~370, numerical  simulations reveal that  substantial variability
can  be  introduced  over  the  lensed image.   This  variability  is,
however, dependent upon the  intrinsic luminosity of the population of
stars under  consideration, with more  luminous populations displaying
more quiescent light curves.

To detect  microlensing induced variability in  the surface brightness
of the giant arcs, a deep monitoring program over a single semester is
required.   Such   a  program  would  allow   easy  identification  of
variability  contaminants,  such   as  novae,  supernovae  and  highly
luminous variable stars, due to their characteristic temporal profiles
(see Lewis \& Ibata 2000).  These contaminating systems are also rare,
while microlensing variability will  appear ubiquitously over the arc.
As  the mass distribution  in the  cluster can  be determined  via the
modeling   of  the  macrolensing   features,  the   identification  of
microlensing variability  will reveal  the MACHO mass  fraction within
the  cluster. Many strongly  lensed systems  are available  for study,
several  displaying very extended  structure [e.g.   cB58 (Yee  et al.
1996)], providing  ideal tools  for the examination  of the  nature of
cluster dark matter.

\section*{Acknowledgments}
GFL  thanks the  computer support  and staff  at  the Anglo-Australian
Observatory  for providing  the  facilities on  which the  simulations
presented in this paper were undertaken.

\newpage

\label{lastpage}


\begin{thebibliography}{99}
%
\bibitem{Abdel1998} 
Abdelsalam, H.M., Saha, P. \& Williams, L.L.R.
1998, \mnras, 294, 734
%
\bibitem[Afonso et al.(1999)]{afonso}
Afonso, C. \etal 
1999, A\&A 344, 63
%
\bibitem{Alcock1999} 
Alcock, C. et al.
1999a, \apj, 521, 602
%
\bibitem{Alcock1999a} 
Alcock, C. et al.
1999b, \apjs, 124, 171
%
\bibitem{Alcock2000} 
Alcock, C.\ et al.
2000a, astro-ph/0001272
%
\bibitem{Alcock2000b} 
Alcock, C.\ et al.
2000b, astro-ph/0002510
%
\bibitem[Ansari et al.(1999)]{Ansari1999} 
Ansari, R.\ et al. 
1999, \aap, 344, L49 
%
\bibitem[Bird, Dickey {\&} Salpeter(1993)]{Bird, et al.1993} 
Bird, C. M., Dickey, J. M. {\&} Salpeter, E. E. 
1993, \apj, 404, 81 
%
\bibitem{Crotts1992}
Crotts, A.P.S.
1992, \apj, 399, L43
%
\bibitem{Fort1994}
Fort, B. \& Mellier, Y.
1994, ARAA, 5, 239
%
\bibitem[Gould(1995)]{Gould1995} 
Gould, A. 
1995, \apj, 455, 44 
%
\bibitem[Henry, Briel {\&} Nulsen(1993)]{Henry, et al.1993} 
Henry, J. P., Briel, U. G. {\&} Nulsen, P. E. J. 
1993, \aap, 271, 413 
%
\bibitem[Hodgkin et al.(2000)]{Hodgkin et al.2000}
Hodgkin, S. T., Oppenheimer, B. R., Hambly, N. C., Jameson, R. F., Smartt,
S. J., Steele, I. A. 
2000, Nature, 403, 57
%
\bibitem{Ibata2000} 
Ibata, R.A., Irwin, M.J., Bienaym\'{e}, O., Scholz, R. \& Guibert, J.
2000, \apj, 532, 41L
%
\bibitem{Jahreis97}
Jahrei\ss\, H. \& Wielen, R.
1997,
in: B. Battrick, M.A.C. Perryman and P.L. Bernacca (eds.):
HIPPARCOS '97. Presentation of the Hipparcos and Tycho catalogues and
first astrophysical results of the Hipparcos space astrometry mission,
Venice, Italy, 13-16 May(1997; ESA SP-402, Noordwijk, p.675-680
%                                                                       
%\bibitem{Kayser1986}
%Kayser, R., Refsdal, S. \& Stabell, R.
%1986, \aap, 166, 36
%
\bibitem[Klypin, Kravtsov, Valenzuela {\&} Prada(1999)]{Kl} 
Klypin, A., Kravtsov, A.\ V., Valenzuela, O.\ {\&} Prada, F. 
1999, \apj, 522, 82
%
\bibitem{Kneib1993} 
Kneib, J.-P., Mellier, Y., Fort, B. \& Mathez, G.
1993, \aap, 273, 367
%
\bibitem[Lasserre et al.(1999)]{}
Lasserre, T. \etal
2000, \aap, 355, L39
%
\bibitem{Lewis2000} 
Lewis, G.F. \& Ibata, R.A.
2000, \apj, {\it In Press}
%
\bibitem{Lewis1995} 
Lewis, G.F. \& Irwin, M.J.
1995, \mnras, 276, 103
%
\bibitem{Lewis1996} 
Lewis, G.F. \& Irwin, M.J. 
1996, \mnras, 283, 225
%
\bibitem{Lewis1993} 
Lewis, G.F., Miralda-Escude, J., Richardson, D.C. \& Wambsganss, J.
1993, \mnras, 261, 647
%
%\bibitem{Schneider1992}
%Schneider, P., Ehlers, J. \& Falco, E. E.
%1992, {\it Gravitational Lenses}, Springer-Verlag Press, Berlin
%
\bibitem[Smail et al.(1996)]{Smail1996} 
Smail, I., Dressler, A., Kneib, J., Ellis, R.\ S., Couch, W.\ J., Sharples, 
R.\ M.\ {\&} Oemler, A.\ J. 
1996, \apj, 469, 508 
%
\bibitem{Soucail1987}
Soucail, G., Fort, B., Mellier, Y. \& Picat, J. P.
1987, A\&A, 172, L14
%
\bibitem{Soucail1988}
Soucail, G., Mellier, Y., Fort, B., Mathez, G. \& Cailloux, M.
1988, \aap, 191, L19
%
\bibitem{Tadros1988} 
Tadros, H., Warren, S.J. \& Hewett, P.C.
1988, New Astr., 42, 11
%
\bibitem{Tadros2000} 
Tadros, H., Warren, S.J. \& Hewett, P.C. 
2000, astro-ph/0003422
%
\bibitem[Walker {\&} Ireland(1995)]{Walker, et al.1995} 
Walker, M. A. {\&} Ireland, P. M. 
1995, \mnras, 275, L41 
%
%\bibitem{Wambsganss1998}
%Wambsganss, J.
%1998, Living Reviews in Relativity, 1, 1
%
\bibitem{Witt1993} 
Witt, H.-J.
1993, \apj, 403, 530
%
\bibitem[Wyithe {\&} Webster(1999)]{Wyithe, et al.1999} 
Wyithe, J. S. B. {\&} Webster, R. L. 
1999, \mnras, 306, 223 
%
\bibitem[Wyithe, Webster {\&} Turner(1999)]{Wyithe, et al.1999b} 
Wyithe, J. S. B., Webster, R. L. {\&} Turner, E. L. 
1999, \mnras, 309, 261 
%
\bibitem[Wyithe, Webster {\&} Turner(2000)]{Wyithe, et al.2000} 
Wyithe, J. S. B., Webster, R. L. {\&} Turner, E. L. 
2000, \mnras, 312, 843 
%
\bibitem{Yee1996}
Yee, H. K. C., Ellingson, E., Bechtold, J., Carlberg, R.G. \& Cuillandre, J.-C.
1996, \aj, 111, 1783
%
\end{thebibliography}
\end{document}